\documentclass[12pt, leqno, a4]{article}
\usepackage{latexsym}
\usepackage{amsmath}
\usepackage{authblk}
\usepackage{amssymb}
\usepackage{amsfonts}
\usepackage{stmaryrd}
\usepackage{color}
\definecolor{Blue}{rgb}{0.,0.,1.}
\definecolor{Red}{rgb}{1.,0.,0.}

\usepackage[T1]{fontenc}
\def\utw{\smash{\rlap{\lower5pt\hbox{$\sim$}}}}
\def\udtw{\smash{\rlap{\lower6pt\hbox{$\approx$}}}}

\def\bbbone{{\mathchoice {\rm 1\mskip-4mu l} {\rm 1\mskip-4mu l}
{\rm 1\mskip-4.5mu l} {\rm 1\mskip-5mu l}}}
\def\bbbc{{\mathchoice {\setbox0=\hbox{$\displaystyle\rm C$}\hbox{\hbox
to0pt{\kern0.4\wd0\vrule height0.9\ht0\hss}\box0}}
{\setbox0=\hbox{$\textstyle\rm C$}\hbox{\hbox
to0pt{\kern0.4\wd0\vrule height0.9\ht0\hss}\box0}}
{\setbox0=\hbox{$\scriptstyle\rm C$}\hbox{\hbox
to0pt{\kern0.4\wd0\vrule height0.9\ht0\hss}\box0}}
{\setbox0=\hbox{$\scriptscriptstyle\rm C$}\hbox{\hbox
to0pt{\kern0.4\wd0\vrule height0.9\ht0\hss}\box0}}}}
\def\bbbe{{\mathchoice {\setbox0=\hbox{\smalletextfont e}\hbox{\raise
0.1\ht0\hbox to0pt{\kern0.4\wd0\vrule width0.3pt
height0.7\ht0\hss}\box0}}
{\setbox0=\hbox{\smalletextfont e}\hbox{\raise
0.1\ht0\hbox to0pt{\kern0.4\wd0\vrule width0.3pt
height0.7\ht0\hss}\box0}}
{\setbox0=\hbox{\smallescriptfont e}\hbox{\raise
0.1\ht0\hbox to0pt{\kern0.5\wd0\vrule width0.2pt
height0.7\ht0\hss}\box0}}
{\setbox0=\hbox{\smallescriptscriptfont e}\hbox{\raise
0.1\ht0\hbox to0pt{\kern0.4\wd0\vrule width0.2pt
height0.7\ht0\hss}\box0}}}}
\def\bbbq{{\mathchoice {\setbox0=\hbox{$\displaystyle\rm Q$}\hbox{\raise
0.15\ht0\hbox to0pt{\kern0.4\wd0\vrule height0.8\ht0\hss}\box0}}
{\setbox0=\hbox{$\textstyle\rm Q$}\hbox{\raise
0.15\ht0\hbox to0pt{\kern0.4\wd0\vrule height0.8\ht0\hss}\box0}}
{\setbox0=\hbox{$\scriptstyle\rm Q$}\hbox{\raise
0.15\ht0\hbox to0pt{\kern0.4\wd0\vrule height0.7\ht0\hss}\box0}}
{\setbox0=\hbox{$\scriptscriptstyle\rm Q$}\hbox{\raise
0.15\ht0\hbox to0pt{\kern0.4\wd0\vrule height0.7\ht0\hss}\box0}}}}
\def\bbbt{{\mathchoice {\setbox0=\hbox{$\displaystyle\rm
T$}\hbox{\hbox to0pt{\kern0.3\wd0\vrule height0.9\ht0\hss}\box0}}
{\setbox0=\hbox{$\textstyle\rm T$}\hbox{\hbox
to0pt{\kern0.3\wd0\vrule height0.9\ht0\hss}\box0}}
{\setbox0=\hbox{$\scriptstyle\rm T$}\hbox{\hbox
to0pt{\kern0.3\wd0\vrule height0.9\ht0\hss}\box0}}
{\setbox0=\hbox{$\scriptscriptstyle\rm T$}\hbox{\hbox
to0pt{\kern0.3\wd0\vrule height0.9\ht0\hss}\box0}}}}
\def\bbbs{{\mathchoice
{\setbox0=\hbox{$\displaystyle     \rm S$}\hbox{\raise0.5\ht0\hbox
to0pt{\kern0.35\wd0\vrule height0.45\ht0\hss}\hbox
to0pt{\kern0.55\wd0\vrule height0.5\ht0\hss}\box0}}
{\setbox0=\hbox{$\textstyle        \rm S$}\hbox{\raise0.5\ht0\hbox
to0pt{\kern0.35\wd0\vrule height0.45\ht0\hss}\hbox
to0pt{\kern0.55\wd0\vrule height0.5\ht0\hss}\box0}}
{\setbox0=\hbox{$\scriptstyle      \rm S$}\hbox{\raise0.5\ht0\hbox
to0pt{\kern0.35\wd0\vrule height0.45\ht0\hss}\raise0.05\ht0\hbox
to0pt{\kern0.5\wd0\vrule height0.45\ht0\hss}\box0}}
{\setbox0=\hbox{$\scriptscriptstyle\rm S$}\hbox{\raise0.5\ht0\hbox
to0pt{\kern0.4\wd0\vrule height0.45\ht0\hss}\raise0.05\ht0\hbox
to0pt{\kern0.55\wd0\vrule height0.45\ht0\hss}\box0}}}}

\def\bbbz{{\mathchoice {\hbox{$\sf\textstyle Z\kern-0.4em Z$}}
{\hbox{$\sf\textstyle Z\kern-0.4em Z$}}
{\hbox{$\sf\scriptstyle Z\kern-0.3em Z$}}
{\hbox{$\sf\scriptscriptstyle Z\kern-0.2em Z$}}}}

\def\diameter{{\ifmmode\oslash\else$\oslash$\fi}}

\newcommand{\HS}{H_\Lambda}
\newcommand{\kak}[1]{(\ref{#1})}

\def\init{\setcounter{equation}{0}}

\newtheorem{theoreme}{Theorem }[section]
\newtheorem{proposition}[theoreme]{Proposition}
\newtheorem{lemma}[theoreme]{Lemma}

\newtheorem{remark}[theoreme]{Remark}

\def\rr{\mathbb{R}}
\def\cc{\mathbb{C}}
\def\nn{\mathbb{N}}

\def\one{\bbbone}

\def\t{{\langle t\rangle}}

\newcounter{smallarabics}
\newenvironment{arabicenumerate}
{\begin{list}{{\normalfont\textrm{(\arabic{smallarabics})}}}
  {\usecounter{smallarabics}\setlength{\itemindent}{0cm}
   \setlength{\leftmargin}{5ex}\setlength{\labelwidth}{4ex}
   \setlength{\topsep}{0.75\parsep}\setlength{\partopsep}{0ex}
   \setlength{\itemsep}{0ex}}}
{\end{list}}

\newcounter{smallroman}

\newcommand{\ben}{\begin{arabicenumerate}}
\newcommand{\een}{\end{arabicenumerate}}

\def\e{{\rm e}}

\def\i{{\rm i}}
\def\d{{\rm d}}
\def\12{\frac{1}{2}}
\def\cinf{C^{\infty}}

\def\proof{\noindent{\bf  Proof. }}

\def\qed{$\Box$}

\def\cH{{\cal H}}

\def\cK{{\cal K}}

\def\K{{\cal K}}

\def\ch{{\mathfrak h}}

\def\p{\partial}
\def\s{{\rm s}}

\def\pfi2{P(\varphi)_{2}}

\def\t{{\scriptscriptstyle\#}}
\newcommand{\beq}{\begin{equation}}
\newcommand{\eeq}{\end{equation}}
\newcommand{\bet}{\begin{theoreme}}
\newcommand{\eet}{\end{theoreme}}
\newcommand{\bel}{\begin{lemma}}
\newcommand{\eel}{\end{lemma}}
\newcommand{\bep}{\begin{proposition}}
\newcommand{\eep}{\end{proposition}}
\newcommand{\bear}[1]{\begin{array}{#1}}
\newcommand{\ear}{\end{array}}

\begin{document}
\def\triple{\interleave}
\def\Gh{\Gamma(\ch)}
\def\Dom{{\rm Dom}}
\def\y{\langle y\rangle}
\def\rx{{\rm x}}
\def\ry{{\rm y}}

\title{Removal of UV cutoff for  the Nelson model with variable coefficients
\protect\footnotetext{AMS 2010 {\it{Subject
Classification}}. 81T10, 81T20, 81Q10, 58C40}
\protect\footnotetext{{\it{Key words and phrases}}. Nelson model,
static space-times, ground state, ultraviolet limit
}}

\author[,1]{C. G\'erard}
\author[,2]{F. Hiroshima}
\author[,3]{A. Panati}
\author[,4]{A. Suzuki}
\affil[1]{D\'epartement de Math\'ematiques, Universit\'e de Paris XI, 91405 Orsay Cedex France}
\affil[2]{Faculty of Mathematics, University of Kyushu, 6-10-1, Hakozaki, Fukuoka, 812-8581, Japan}
\affil[3]{UMR6207 Universit\'e Toulon-Var
83957 La Garde Cedex France}
\affil[4]{
Department of Mathematics, Faculty of Engineering, Shinshu University, 4-17-1 Wakasato, Nagano 380-8553, Japan}

\date{\today}

\maketitle
\begin{abstract}
We consider  the Nelson model with variable coefficients.
Nelson models with variable
coefficients arise when one replaces in the usual Nelson model
the flat Minkowski metric by a  static metric, allowing also the
boson mass to depend on position.
We  study the removal of the ultraviolet cutoff.
\end{abstract}

\section{The Nelson Hamiltonian with variable coefficients}\init\label{sec2}
\subsection{Introduction}
\label{intro.4}
The Nelson model \cite{Ne} describes a spinless nonrelativistic particle linearly coupled to a scalar bose field.
After adding an ultraviolet (UV) cutoff, this model can be defined as a self-adjoint operator on some Hilbert space.
In [Ne], E. Nelson was able to remove the UV cutoff and to define the Hamiltonian as a self-adjoint operator without UV cutoff on the original Hilbert space.

We extend the Nelson model to the case with variable coefficients, which realizes the Nelson model defined on a static Lorentzian manifold. 
In a series  of papers \cite{GHPS1,GHPS2,GHPS3}, we
show the existence or absence of ground states of the variable coefficients Nelson model $H(\rho)$ with a certain UV cutoff $\rho$.
In this paper we  consider the removal of UV cutoff
 for variable coefficients Nelson models.
 Denoting by $H^{\kappa}$ the Nelson Hamiltonian $H(\rho^\kappa)$ for the cutoff function $\rho^{\kappa}(\rx)=\kappa^{3}\rho(\kappa \rx)$, we construct a  particle potential
$E^{\kappa}(X)$ such that $H^{\kappa}- E^{\kappa}(X)$ converge in strong resolvent sense to a bounded below selfadjoint operator $H^{\infty}$.
The removal of the UV cutoff involves as in the constant coefficients case a sequence of unitary dressing operators $U^{\kappa}$. In contrary to the constant coefficients case, where all computations can be conveniently done in momentum space (after conjugation by Fourier transform), we have
to use instead {\em pseudodifferential calculus}. Some of the rather advanced facts on pseudodifferential calculus which we will need are recalled in Appendix \ref{app1}.

\subsection{Notation}
We collect here some notation for
reader's convenience.

We
denote
 by $x\in \rr^{3}$ (resp. $X\in \rr^{3}$) the boson (resp. electron) position.
As usual we set $D_{x}= \i^{-1}\nabla_{x}$, $D_{X}= \i^{-1}\nabla_{X}$.
If $x\in \rr^{d}$, we set
$\langle x\rangle= (1+ |x|^{2})^{\12}$.
The domain of a linear operator $A$ on some Hilbert space  will be denoted by $\Dom A$, and its spectrum by $\sigma(A)$.
If $\ch$ is a Hilbert space, the {\em bosonic Fock space} over $\ch$ denoted by $\Gamma_{\rm s}(\ch)$ is
\[
\Gamma_{\rm s}(\ch) =\bigoplus_{n=0}^{\infty}\otimes_{\rm s}^{n}\ch.
\]
We denote by  $a^{*}(h)$ and $a(h)$ for $h\in \ch$ the creation operator and the annihilation operator, respectively,  which acts on $\Gamma_{\rm s}(\ch)$. The (Segal) {\em field operators}
$\phi(h)$ are defined as \beq
\phi(h) =\frac{1}{\sqrt{2}}(a^{*}(h)+ a(h)).
\eeq
If $\cK$ is another Hilbert space and $v\in B(\cK, \cK\otimes\ch)$,
one defines  the operators $a^{*}(v)$ and
$a(v)$ as unbounded operators on $\cK\otimes \Gamma_{\rm s}(\ch)$ by
\[\begin{array}{l}
a^*(v)\Big|_{\cK\otimes\bigotimes_\s^n\ch}
 =\sqrt{n+1}\Big(\one_\cK\otimes {\cal S}_{n+1}\Big)
\Big(v\otimes\one_{\bigotimes_\s^n\ch}\Big),\\[3mm]
a(v) =\big(a^*(v)\big)^*,\\[3mm]
\displaystyle
\phi(v) =\frac{1}{\sqrt2}(a^\ast(v)+a(v)).
\end{array}\]
If $b$ is a selfadjoint operator on $\ch$ its second quantization $\d \Gamma(b)$ is defined as
 \[
\d\Gamma(b)\Big|_{\bigotimes_\s^n\ch}
=\sum\limits_{j=1}^n\underbrace{\one\otimes\cdots\otimes\one}_{j-1}
\otimes b\otimes \underbrace{\one\otimes\cdots\otimes\one}_{n-j}.
\]
The number operator $N$ is defined by the second quantization of the identity operator $\one$:
$$N= \d\Gamma(\one).$$
The annihilation opeator and the creation operator satisfy the estimate:
\beq
\|a^{\sharp}(v)(N+1)^{-\12}\|\leq \|v\|,
\label{ju88}
\eeq
where $\|v\|$ is the norm of $v$ in $B(\K,\K\otimes\ch)$.

\subsection{Field Hamiltonian}\label{sec2.2}
Let
\begin{align*}
h_{0} =&-\sum_{1\leq j,k\leq d} c(x)^{-1}\p_{j}a^{jk}(x)\p_{k}c(x)^{-1},\\[2mm]
 h =& h_{0}+ m^{2}(x),
\end{align*}
with $a^{jk}$, $c$, $m$  are real functions and
\[
{\rm (B)} \ \begin{array}{l}
C_{0}\one\leq [a^{jk}(x)]\leq C_{1}\one,  \ C_{0}\leq c(x)\leq C_{1}, \ C_{0}>0, \\[2mm]
\p_{x}^{\alpha}a^{jk}(x)\in O(\langle x\rangle^{-1}), \ |\alpha|\leq 1,\\[2mm]
 \p_{x}^{\alpha}c(x)\in O(1), \ |\alpha|\leq 2,\\[2mm]
\p_{x}^{\alpha}m(x)\in O(1), \ |\alpha|\leq 1.
\end{array}
\]
Clearly $h$ is selfadjoint on  $H^{2}(\rr^{3})$ and
$h\geq 0$. The one-particle space is
given by
\[
\ch =L^{2}(\rr^{3}, \d x)
\]
and
one-particle energy by
the selfadjoint operator:
\[\omega =h^{\12}.
\]
It
can be easily seen that
\ben
\item ${\rm Ker}\omega=\{0\}$.
\item
Assume in addition to  (B) that $\lim_{x\to \infty}m(x)=0$. Then $\inf \sigma(\omega)=0$.
\een
The field Hamiltonian is
\[
\d\Gamma(\omega),
\]
acting on the bosonic Fock space $\Gamma_{\rm s}(\ch)$.

\subsection{Electron Hamiltonian}\label{sec2.1}
We define the electron Hamiltonian as
\[
K = K_{0}+ W(X),
\]
where
\[
 K_{0}=\sum_{1\leq j, k\leq 3}D_{X_{j}} A^{jk}(X)D_{X_{k}},
\]
 acting on $\cK = L^{2}(\rr^{3},\d X)$,
 and
\[
{\rm (E)}\  \ C_{0}\one \leq [A^{jk}(X)]\leq C_{1}\one,\ C_{0}>0.
\]
We assume that $W(X)$ is a real potential such that $K_{0}+W$ is
essentially selfadjoint and bounded below. We denote by $K$ the
closure of $K_{0}+W$.
\subsection{Nelson Hamiltonian with variable coefficients}\label{sec2.3}
The constant
\[
m=\inf\sigma(\omega)\geq 0
\]
can be viewed as the {\em mass} of the scalar bosons.
The Nelson
Hamiltonian defined  below will be called {\em massive} (resp. {\em
massless}) if $m>0$ (resp. $m=0$).
Let $\rho\in S(\rr^{3})$, with $\rho\geq 0$, $q=\int_{\rr^{3}} \rho(y)\d y\neq 0$.  We set
\[
 \rho_{X}(x)= \rho(x-X)
\]
  and define the  UV cutoff scalar bose fields as
\beq\label{e0.1}
\varphi_{\rho}(X) = \phi( \omega^{-\12} \rho_{X}),
\eeq
where $\phi(f)$ is the Segal field operator.
The {\em Nelson Hamiltonian} with UV cutoff $\rho$ is
given by
\beq\label{e2.1}
H(\rho) = K\otimes\one + \one \otimes \d\Gamma(\omega)+ \varphi_{\rho}(X),
\eeq
acting on
the Hilbert space:
\[
\cH= \cK\otimes\Gamma_{\rm s}(\ch).
\]
Set also
\[
H_{0} =K\otimes\one + \one \otimes \d\Gamma(\omega),
\]
which is selfadjoint on its natural domain.  Moreover assume  hypotheses (E) and  (B). Then $H$ is selfadjoint and bounded below on $D(H_{0})$.

\section{Removal of the UV cutoff}\label{sec3}\init
\subsection{Nelson Hamiltonians with constant coefficients
}
In \cite{Ne} Nelson considered the limit of $H(\rho)$  for
\begin{align}
\label{n1}
\omega&= \omega(D_x)= (-\Delta_x+ m^2)^{\12}
,\quad m\geq 0,\\
\label{n2}
  K&=-\12 \Delta_X+W(X),
  \end{align}
 when $\rho$ tends to the Dirac mass $\delta$, equivalently, the Fourier transform $\hat\rho$ to $(2\pi)^{-3/2}$.

We quickly review the results in \cite{Ne}.
 In the rest of this subsection we take the momentum representation for the field variables.
  Let $H_{\rm Stand}
  $ be the constant coefficients  Nelson model defined by $H(\rho)$ with \kak{n1} and \kak{n2}.
Suppose $m=0$
 and $\hat \rho_\Lambda(k)=1$ for $|k|<\Lambda$ and $\hat\rho_\Lambda(k)=0$ otherwise.
We denote by $H_\Lambda$ the Hamiltonian $H_{\rm Stand}$ with
 $\hat\rho$ replaced by $\hat\rho_\Lambda$.
Let
\beq\label{n7}
\hat T (k)=\omega(k)+|k|^2/2
\eeq and
\beq\label{n6}
E_\Lambda=-\12\int_{{\mathbb R}^3}
{\omega(k)^{-1}}
\chi(\omega(k)>\sigma)\hat T (k)^{-1}
|\hat\rho_\Lambda|(k)^2dk,
\eeq
where $\sigma>0$ is arbitrary and $\chi(\omega(k)>\sigma)$ is an IR cutoff function given by
$\chi(\omega(k)>\sigma)=0$ for $\omega(k)<\sigma$ and $1$ for $\omega(k)\geq \sigma$.
Define the dressing transformation by
\beq\label{n3}
U_\Lambda=e^{i\phi(i\beta_X)},
\eeq
where
\beq\label{n4}
\beta_X(k)=
-\hat T (k)^{-1} \chi(\omega(k)>\sigma)\omega(k)^{-\12}
e^{-ik X}\hat \rho_\Lambda(k).
\eeq
It is easy to see that $U_\Lambda\to U_\infty$ strongly as $\Lambda\to\infty$, where $U_\infty$ is given by $U_\Lambda$ with $\hat\rho_\Lambda$ replaced by $1$. Instead of $\HS$, Nelson considers the dressing-transformed Hamiltonian:
\beq
\label{n5}
\widehat \HS=U_\Lambda \HS U_\Lambda^\ast.
\eeq
 \begin{proposition}
{\rm \cite{Ne}}
There exists a bounded below self-adjoint operator $H_\infty$ such that
$\widehat \HS-E_\Lambda$ converges to $H_\infty$ as $\Lambda\to\infty$ in
the uniform resolvent sense, and
$
\HS-E_\Lambda$
to
$U_\infty^\ast H_\infty U_\infty$
in the strong resolvent sense.
\end{proposition}
\begin{remark}
{\rm 
Nelson \cite{Ne} actually considered only the case of $m>0$. It can be however extended to the case of $m=0$.
}
\end{remark}

In this section we study the same problem for the Nelson model with variable coefficients.

\subsection{Preparations}\label{sec3.2}
In the constant coefficients Nelson model,   the one-particle operator $\omega$ is diagonalized using the Fourier transform. In the variable coefficients Nelson Hamiltonian
 we will use instead the pseudodifferential calculus to define operators and constants corresponding to  \kak{n6}, \kak{n3}, \kak{n4} and  \kak{n5}.  In particular
the renormalization constant $E_\Lambda$ will be changed
to a function $E(X)$.

We denote by $S^{0}(\rr^{3})$ the space
\[
S^{0}(\rr^{3})=\{f\in C^{\infty}(\rr^{3})\  | \  |\p_{x}^{\alpha}f(x)|\leq C_{\alpha}, \ \alpha\in \nn^{3} \}.
\]
We will assume in addition to hypotheses (E) and  (B) that
\[
{\rm (N)} \ A_{jk}(X), \ a_{jk}(x), \ c(x),\  m^{2}(x)\in S^{0}(\rr^{3}).
\]
It is easy to see that $h$ can be rewritten as
\[
h= \sum_{jk}D_{j}c^{-2}(x)a^{jk}(x)D_{j}+ v(x),
\]
where $v\in S^{0}(\rr^{3})$, and that $c^{-2}(x)a^{jk}(x)\in S^{0}(\rr^{3})$. Changing notation, we will henceforth assume that
\[
h= \sum_{jk}D_{j}a^{jk}(x)D_{j}+ v(x),
\]
where $[a_{jk}](x)$ satisfies (B) and $a^{jk}, \ v\in S^{0}(\rr^{3})$.
 We refer the reader to Appendix
\ref{app1} for the notation and for some background on pseudodifferential calculus.
It will be useful later to consider $\omega= h^{\12}$ as a
pseudodifferential operator. Note first that
\[
h= h^{\rm w}(x, D_{x}),
\]
for
\[
h(x,\xi)= \sum_{1\leq j,k\leq 3 }\xi_{j}a^{jk}(x)\xi_{k}+ c(x).
\]
The symbol $h(x,\xi)$ belongs to $S(\langle \xi\rangle^{2}, g)$, for
the standard metric
$$g= \d x^{2}+ \langle \xi\rangle^{-2}\d \xi^{2},$$
and is elliptic in this class.  By Lemma \ref{a.1} and Theorem  \ref{a.3},
we know that if   $f\in S^{p}(\rr )$,
then the operator $f(h)$ belongs to $\Psi^{\rm w}(\langle
\xi\rangle^{2p}, g)$.

If the model is massive, then picking a function $f\in S^{\12}(\rr)$
 equal to $\lambda^{\12}$ in $\{\lambda\geq m/2\}$, we
see that  $\omega= f(h)\in \Psi^{\rm w}(\langle \xi\rangle,
g)$.
If the model is massless, we   fix $\sigma>0$ ($\sigma= 1$ will do)
and pick $f\in \cinf(\rr)$ such that
\[
f(\lambda)=\left\{
\begin{array}{ll}
\lambda^{\12} & \hbox{ if }|\lambda|\geq 4\sigma^{2}, \\
\sigma& \hbox{ if } |\lambda|\leq \sigma^2.
\end{array}
\right.
\]
 We set
\[
\omega_{\sigma} = f(h).
\]
Again by Theorem  \ref{a.3} we know that $\omega_{\sigma}$ belongs
to $\Psi^{\rm w}(\langle \xi\rangle, g)$.
In the massive case $\omega= f(h)$ will also be denoted by $\omega_\sigma$.
Consider now the operator
\[
T = K_{0}\otimes \one + \one\otimes \omega_{\sigma},
\]
acting on $L^{2}(\rr^{3}, \d X)\otimes L^{2}(\rr^{3}, \d x)$. Clearly
$T$ is selfadjoint on its natural domain and $T\geq \sigma$.

\begin{lemma}\label{3.1}
Set
\[
M(\Xi, \xi) = \langle \Xi\rangle^{2}+ \langle \xi\rangle, \ G=
\d X^{2}+ \d x^{2}+  \langle\Xi\rangle^{-2}\d \Xi^{2}+ \langle
\xi\rangle^{-2}\d \xi^{2}.
\]
Then $T^{-1}$ belongs to $\Psi^{\rm w}(M^{-1}, G)$.
\end{lemma}
\proof By Lemma \ref{a.2}, the metric $G$ and weight $M$ satisfy
all the conditions in Subsect. \ref{app1.1}. Clearly $T\in \Psi^{\rm w}(M,  G)$.
We pick a function $f\in
S^{-1}(\rr)$ such that $f(\lambda)= \lambda^{-1}$ in $\{\lambda\geq
\sigma/2\}$. By Theorem  \ref{a.3} $T^{-1}= f(T)\in \Psi^{\rm w}(
M^{-1}, G)$. \qed

Let us fix another  cutoff function $F(\lambda\geq \sigma)\in \cinf(\rr)$ with
\[
F(\lambda\geq \sigma)=\left\{
\begin{array}{l}
1\hbox{ for }|\lambda|\geq 4\sigma, \\
0\hbox{ for }|\lambda|\leq 2\sigma,
\end{array}
\right.
\]
and set
\[
F(\lambda\leq \sigma) = 1-F(\lambda\geq \sigma).
\]
\begin{lemma}\label{3.2}
Set
\[
\beta(X, x)= \beta_{X}(x) = -T^{-1}
F(\omega\geq\sigma)\omega^{-\12}\rho_{X}=-T^{-1}
F(\omega\geq\sigma)\omega_{\sigma}^{-\12}\rho_{X}.
\]
Then
\ben
\item $\beta\in \cinf(\rr^{6})$.
\item Let $0\leq \alpha<1$. Then  $\omega^{\alpha}\beta_{X}\in
L^{2}(\rr^{3}, \d x)$ and there exists $s>3/2$ such that
\[
\|\omega^{\alpha}\beta_{X}\|_{L^{2}(\rr^{3}, \d x)}\leq
C\|\rho\|_{H^{-s}(\rr^{3})},
\]
uniformly in $X$.
\item Let $\alpha>0$. Then  $\omega^{-\alpha}\nabla_{X}\beta_{X}\in
L^{2}(\rr^{3}, \d x)$ and there exists $s>3/2$ such that
\[
\|\omega^{-\alpha}\nabla_{X}\beta_{X}\|_{L^{2}(\rr^{3}, \d x)}\leq
C\|\rho\|_{H^{-s}(\rr^{3})},
\]
uniformly in $X$.
\item One has
\[
\omega^{-\12}\rho_{X}+ (K_{0}\otimes \one + \one \otimes
\omega)\beta_{X}= \omega^{-\12}F(\omega\leq \sigma)\rho_{X}.
\]
\een
\end{lemma}
\proof The function $\rho_{X}(x)$ is clearly $C^{\infty}$ in $(X, x)$, so (1)
follows from the fact that $T^{-1}$ and $\omega_{\sigma}^{-\12}F(\omega\geq\sigma)$
are pseudodifferential operators.

 We claim that there exists a symbol
$ b_{X}(x, \xi)=b(X, x,\xi)$  such that
\beq\label{e3.4}
\begin{array}{l}
b(X, x, \xi)\in S(\langle \xi\rangle^{-5/2}, \d X^{2}+ \d x^{2}+
\langle \xi\rangle^{-2}\d\xi^{2}),\\[2mm]
\beta_{X}= b_{X}^{(1, 0)}(x, D_{x})\rho_{X}.
\end{array}
\eeq

Let us prove our claim.
Let
 $B(X, x, \Xi, \xi)\in S(M^{-1}, G)$ be  the $(1, 0)$ symbol of $T^{-1}$.
Applying Lemma \ref{3.1} and (\ref{e.a6bis}), we know that $T^{-1}\in
\Psi^{(1, 0)}(M^{-1}, G)$. Setting $w(X, x)= T^{-1}\rho_{X}$, this
yields
\beq
\begin{array}{rl}
w(X, x)=&(2\pi)^{-3}\int \e^{\i (X\cdot \Xi+ x\cdot \xi)} B(X, x, \Xi,
\xi) \delta( \xi+ \Xi)\hat{\rho}(\xi)\d \xi \d \Xi\\[2mm]
=&(2\pi)^{-3}\int\e^{\i (x-X)\cdot \xi}B(X, x, -\xi,
\xi)\hat{\rho}(\xi)\d \xi\\[2mm]
=& b_{X}^{(1, 0)}(x, D_{x})\rho_{X}
\end{array}
\label{e3.1}
\eeq
for
\beq\label{arlat}
b_{X}(x, \xi)= B(X, x, -\xi, \xi).
\eeq
This implies that
\[
b_{X}\in S(\langle \xi\rangle^{-2}, \d X^{2}+ \d x^{2}+
\langle \xi\rangle^{-2}\d\xi^{2}).
\]
Applying once again Theorem  \ref{a.3}, we know that
$F(\omega\geq\sigma)\omega_{\sigma}^{-\12}\in \Psi^{(1, 0)}(\langle\xi\rangle^{-\12},
g)$.  By the composition property  (\ref{e.a6ter}), we obtain our
claim.

 (2) follows from (\ref{e3.4}), if we note that
$\omega^{\alpha}F(\omega\geq\sigma)\omega_{\sigma}^{-\12}\in \Psi^{(1, 0)}(\langle
\xi\rangle^{\alpha-\12}, g)$ and use the mapping property of
pseudodifferential operators between Sobolev spaces recalled in
(\ref{e.app00}). (3) is proved similarly, using that
\[
\nabla_{X}b_{X}(x, D_{x})\rho_{X}= \p_{X}b_{X}(x, D_{x})\rho_{X}-
b_{X}(x, D_{x})\nabla_{x}\rho_{X}.
\]
Finally (4) follows from the fact that $(\omega-
\omega_{\sigma})F(\omega\geq \sigma)=0$. \qed

\subsection{Dressing transformation}\label{sec3.3}
Let $\rho$ be a charge density as above. We set for $\kappa\gg 1$
\[
\rho^{\kappa}(x) = \kappa^{3}\rho(\kappa x), \ \rho^{\kappa}_{X}(x)=
\rho^{\kappa}(x-X),
\]
so that
\beq\label{e3.5b}
\lim_{\kappa\to \infty}\rho^{\kappa}_{X}= q\delta_{X} \hbox{ in
}H^{-s}(\rr^{3}), \ \forall \ s>3/2,
\eeq
where $q=\int_{\mathbb R^3} \rho(y) dy$.
This implies
\beq
\|\rho^{\kappa}_{X}\|_{H^{-s}(\rr^{3})}\leq C, \hbox{ uniformly in }X,
\kappa, \hbox{ for all }s>3/2.
\label{e3.5}
\eeq

We set\[
H^\kappa= H(\rho^{\kappa}),
\]   and as in \cite{Ne}
\[
 U^{\kappa} = \e^{\i \phi(\i \beta^{\kappa}_{X})},
\]
which is a unitary operator on $\cH$. (Recall that
$\beta^{\kappa}_{X}$ is defined in Lemma \ref{3.2}).

\begin{proposition}\label{3.3}
Set
\[
\begin{array}{rl}
 a^{\kappa}_{j}(X)=& \frac{1}{\sqrt{2}}
a(\nabla_{X_{j}}\beta^{\kappa}_{X}),\\[2mm]
R^{\kappa}=& 2\sum_{j, k} \nabla_{X_{j}}A_{jk}(X)a^{\kappa}_{k}(X)-
 a^{\kappa*}_{j}(X)A_{jk}(X)\nabla_{X_{k}}\\[2mm]
&+\sum_{j,k} 2a^{\kappa*}_{j}(X)A_{jk}(X)a^{\kappa}_{k}(X)-
a^{\kappa*}_{j}(X)A_{jk}(X)a^{\kappa*}_{k}(X) -
a^{\kappa}_{j}(X)A_{jk}(X)a^{\kappa}_{k}(X),\\[2mm]
V^{\kappa}(X)=&-(\rho^{\kappa}_{X}| \omega^{-1}F(\omega\geq \sigma)T^{-1}\rho^{\kappa}_{X}) + \12 (T^{-1}\rho^{\kappa}_{X}| F^{2}(\omega\geq \sigma) T^{-1}\rho^{\kappa}_{X})\\[2mm]
&+\12 \sum_{jk}A_{jk}(X)(\nabla_{X_{j}}T^{-1}\rho^{\kappa}_{X}|\omega^{-1}F^{2}(\omega\geq \sigma)\nabla_{X_{k}}T^{-1}\rho^{\kappa}_{X}).
\end{array}
\]
Then
\[
\begin{array}{rl}
 U^{\kappa}H^{\kappa}U^{\kappa*}=& K +\d\Gamma(\omega)+
\phi(\omega^{-\12}F(\omega\leq \sigma)\rho^{\kappa}_{X})\\[2mm]
&+ R^{\kappa} + V^{\kappa}(X).
\end{array}
\]
\end{proposition}
\proof We recall some well-known
identities
\beq\label{urlu}
 U^{\kappa} (\d\Gamma(\omega)+ \phi(\omega^{-\12}\rho_{\kappa,
X}))U^{\kappa*}
= \d\Gamma(\omega)+ \phi( \omega \beta^{\kappa}_{X}+
\omega^{-\12}\rho^{\kappa}_{X})+ {\rm Re}(
\frac{\omega}{2}\beta^{\kappa}_{X} + \omega^{-\12}\rho^{\kappa}_{X}|
\beta^{\kappa}_{X}).
\eeq
Note that the scalar product in the rhs is real valued, since  $\rho_{X}^{\kappa}$, $\beta^{\kappa}_{X}$ and  $\omega$ are real vectors
and operators.  Using once more that $\beta^{\kappa}_{X}$ is real, we see that the operators $\phi(\i \beta^{\kappa}_{X})$
for different  $X$ commute, which yields
\[
U_{\kappa}D_{X_{j}}U^{\kappa*}= D_{X_{j}}- \phi(\i
\nabla_{X_{j}}\beta^{\kappa}_{X}),
\]
and hence
\[
U^{\kappa} KU^{\kappa*}=\sum_{j,k} \left(D_{X_{j}}- \phi(\i
\nabla_{X_{j}}\beta^{\kappa}_{X})\right)A_{jk}(X)\left(D_{X_{k}}- \phi(\i
\nabla_{X_{k}}\beta^{\kappa}_{X})\right)+ W(X).
\]
We expand the squares in the r.h.s. using the definition of $a_{j}^{\kappa}(X)$ in the proposition. After rearranging the various terms, we obtain
\[
\begin{array}{rl}
U^{\kappa} KU^{\kappa*}=&K + \phi(K_{0} \beta^{\kappa}_{X})\\[2mm]
&+ 2\sum_{j, k} \nabla_{X_{j}}A_{jk}(X)a^{\kappa}_{k}(X)-
a^{\kappa*}_{j}(X)A_{jk}(X)\nabla_{X_{k}}\\[2mm]
&+\sum_{j,k} 2
a^{\kappa*}_{j(X)}A_{jk}(X)a^{\kappa}_{k}(X)-a^{\kappa*}_{j}(X)A_{jk}(X)
a^{\kappa*}_{k}(X) - a^{\kappa}_{j}(X)A_{jk}(X)a^{\kappa}_{k}(X)\\[2mm]
&+\12\sum_{jk}A_{jk}(X)(\nabla_{X_{j}} \beta^{\kappa}_{X}| \nabla_{X_{k}}\beta^{\kappa}_{X}).
\end{array}
\]
This yields
\[
\begin{array}{rl}
U^{\kappa}H^\kappa U^{\kappa*}=& K+ \d\Gamma(\omega) \\[2mm]
&+ 2\sum_{j, k} \nabla_{X_{j}}A_{jk}(X)a^{\kappa}_{k}(X)-
 a^{\kappa*}_{j}(X)A_{jk}(X)\nabla_{X_{k}}\\[2mm]
&+\sum_{j,k} 2
a^{\kappa*}_{j}(X)A_{jk}(X)a^{\kappa}_{k}(X)-a^{\kappa*}_{j}(X)A_{jk}(X)
a^{\kappa*}_{k}(X) - a^{\kappa}_{j}(X)A_{jk}(X)a^{\kappa}_{k}(X)\\[2mm]
&+ \phi(\omega^{-\12} \rho^{\kappa}_{X}+ (K_{0}+ \omega) \beta^{\kappa}_{X})\\[2mm]
&+ (\omega^{-\12} \rho^{\kappa}_{X}+ \12\omega \beta^{\kappa}_{X}|
\beta^{\kappa}_{X})+ \12\sum_{jk}A_{jk}(X)(\nabla_{X_{j}} \beta^{\kappa}_{X}| \nabla_{X_{k}}\beta^{\kappa}_{X}).
\end{array}
\]
The sum of the second and third lines equals $R^{\kappa}$. By Lemma
\ref{3.2}, the fourth line equals $\phi(\omega^{-\12}F(\omega\leq \sigma) \rho_{X})$.   The fifth line equals $V^{\kappa}(X)$, using the definition of  $\beta_{X}$.
\qed
\subsection{Removal of the ultraviolet cutoff}\label{sec3.4}
Set
\[
h_{0}(x, \xi)= \sum_{1\leq j, k\leq 3}\xi_{j}a_{jk}(x)\xi_{k},
\ K(X, \xi)= \sum_{1\leq j, k\leq 3}\xi_{j}A_{jk}(X)\xi_{k}.
\]
and
\beq\label{e3.01}
E^{\kappa}(X) = -\12(2\pi)^{-3}\int(h_{0}(X, \xi)+1)^{-\12}K(X, \xi)(K(X, \xi)+
1)^{-2}|\hat{\rho}|^{2}(\xi\kappa^{-1})\d \xi.
\eeq
\begin{lemma}\label{3.4}

Then there exists a bounded continuous potential $V_{\rm ren}$ such
that
\[
\lim_{\kappa\to +\infty} V^{\kappa}(X)
-E^{\kappa}(X)= V_{\rm ren}(X),
\]
in $L^{\infty}(\rr^{3})$.
\end{lemma}
We will prove this lemma later. 
We are in the position to state the main theorem.

\bet\label{3.7} Assume hypotheses (E), (B), (N). Then
the family of selfadjoint operators
\[
H^{\kappa}- E^{\kappa}(X)
\]
converges in strong resolvent sense to a bounded below selfadjoint
operator $H^{\infty}$.
\eet
\proof By Prop. \ref{3.8} below, $U^{\kappa}(H^{\kappa}-
E^{\kappa}(X))U^{\kappa*}$ converges in norm  resolvent sense  to
$\hat{H}^{\infty}$. Moreover by Lemma \ref{3.2} (2),
$\beta^{\kappa}_{X}$ converges in $B(\cK, \cK\otimes\ch)$ when
$\kappa\to \infty$, hence $U^{\kappa}$ converges strongly to some
unitary operator $U^{\infty}$. It follows that $H^{\kappa}$ converges
in strong resolvent sense to
\[
H^{\infty}= U^{\infty*}\hat{H}^{\infty}U^{\infty}. \ \Box
\]

{\bf Proof of Lemma \ref{3.4}.}
For simplicity we will assume that the model is massive ($m>0$), which allows to remove the cutoffs $F(\omega\geq \sigma)$ in the various formulas. The massless case can be treated similarly. Recall that
\begin{equation}
\label{urlit}\begin{array}{rl}
T^{-1}\rho^{\kappa}_{X}&=b_{X}(x, D_{x})\rho^{\kappa}_{X},\\[2mm]
\p_{X}T^{-1}\rho^{\kappa}_{X}&= \p_{X}b_{X}(x, D_{x})\rho^{\kappa}_{X}- b_{X}(x, D_{x})\p_{x}\rho^{\kappa}_{X},
\end{array}
\end{equation}
where $b_{X}(x,\xi)$ is defined in (\ref{arlat}).  Plugging the second identity in (\ref{urlit}) into the formula giving $V_{\kappa}(X)$ we get
\[
V^{\kappa}(X)= V_{1}^{\kappa}(X)+ V_{2}^{\kappa}(X),
\]
for
\[
\begin{array}{rl}
V_{1}^{\kappa}(X)=&\12  \|b_{X}(x, D_{x})\rho^{\kappa}_{X}\|^{2}+ \12\sum_{jk}A_{jk}(X)(\p_{X_{j}}b_{X}(x, D_{x})\rho^{\kappa}_{X}|\omega^{-1}\p_{X_{k}}b_{X}(x, D_{x})\rho^{\kappa}_{X})\\[2mm]
&-\sum_{jk}A_{jk}(X)(\p_{X_{j}}b_{X}(x, D_{x})\rho^{\kappa}_{X}|\omega^{-1}b_{X}(x, D_{x})\p_{x_{k}}\rho^{\kappa}_{X}),\\[2mm]
V_{2}^{\kappa}(X)=&-(\rho^{\kappa}_{X}| \omega^{-1}b_{X}(x, D_{x})\rho^{\kappa}_{X})+\12\sum_{jk}A_{jk}(X)(b_{X}(x, D_{x})\p_{x_{j}}\rho^{\kappa}_{X}| \omega^{-1}b_{X}(x, D_{x})\p_{x_{k}}\rho^{\kappa}_{X}).
\end{array}
\]
We will use that
\begin{equation}
\label{urlot}
\begin{array}{l}
\rho^{\kappa}_{X}\to q\delta_{X}\hbox{ in }H^{s}(\rr^{3}), \ \forall s<-\frac{3}{2}, \\[2mm]
\p_{x}\rho^{\kappa}_{X}\to q\p_{x}\delta_{X}\hbox{ in }H^{s}(\rr^{3}), \ \forall s<-\frac{5}{2},\hbox{ uniformly in }X\in \rr^{3},
\end{array}
\end{equation}
where we recall that $q= \int_{\rr^{3}} \rho(y)\d y$. Using that  $b_{X}(x, \xi)\in S(\langle \xi\rangle^{-2} , g)$ and the mapping properties of pseudodifferential operators between Sobolev spaces,  we obtain that
\[
\lim_{\kappa\to \infty}V_{1}^{\kappa}(X)=  V_{1}^{\infty}(X)\hbox{ exists uniformly for }X\in \rr^{3},
\]
and $V_{1}^{\infty}(X)$ is a bounded continuous function, whose exact expression is obtained by replacing $\rho^{\kappa}_{X}$ by $q\delta_{X}$ in the formula giving $V_{1}^{\kappa}(X)$.

We now consider the potential $V_{2}^{\kappa}(X)$, which will be seen to be logarithmically divergent when $\kappa\to \infty$. To extract its divergent part, we use symbolic calculus.    We will use only the $(1, 0)$ quantization and omit the corresponding superscript.
We first use Prop. \ref{a.4} for the metric $G$
defined in Lemma \ref{3.1}. Note that the `Planck constant' for the
metric $G$ is
\[
\lambda(X, x, \Xi, \xi)= \min(\langle \Xi\rangle, \langle \xi\rangle).
\]
Applying Prop. \ref{a.4}, we obtain that the  symbol  $b_{X}(x, \xi)$ in
(\ref{e3.1}) equals
\beq\label{e3.6}
\begin{array}{rl}
b_{X}(x, \xi)=& (K(X, \xi)+ (h_{0}(x, \xi)+1)^{\12})^{-1}+ S(\langle
\xi\rangle^{-3}, g)\\[2mm]
=&(K(X, \xi)+ 1)^{-1}+ S(\langle
\xi\rangle^{-3}, g).
\end{array}
\eeq
The same argument for the metric $g$ shows that $\omega^{-1}=
d(x, D_{x})$ for
\beq\label{e3.7}
d(x, \xi)= (h_{0}(x, \xi)+ 1)^{-\12}+ S(\langle \xi\rangle^{-2}, g).
\eeq
Combining (\ref{e3.6}) and (\ref{e3.7}) we get that
\begin{equation}
\label{arlo}
\begin{array}{l}
\omega^{-1}b_{X}(x, D_{x})= c_{X}(x, D_{x})+ r_{X}(x, D_{x}), \\[2mm]
b_{X}^{*}(x, D_{x})\omega^{-1}b_{X}(x, D_{x})= d_{X}(x, D_{x})+ s_{X}(x, D_{x}),
\end{array}
\end{equation}
where
\begin{equation}
\label{alr1}
\begin{array}{l}
c_{X}(x, \xi)= (h_{0}(x, \xi)+1)^{-\12}(K(X, \xi)+1)^{-1},\\[2mm]
d_{X}(x, \xi)= (h_{0}(x, \xi)+1)^{-\12}(K(X, \xi)+1)^{-2},\\[2mm]
r_{X}(x, \xi)\in S(\langle \xi\rangle^{-4}, g), \ s_{X}(x, \xi)\in S(\langle \xi\rangle^{-6}, g),\hbox{ uniformly in }X\in \rr^{3}.
\end{array}
\end{equation}
Setting
\[
\tilde{V}_{2}^{\kappa}(X)= -(\rho^{\kappa}_{X}| c_{X}(x, D_{x})\rho^{\kappa}_{X})+\12\sum_{jk}A_{jk}(X)(\p_{x_{j}}\rho^{\kappa}_{X}| d_{X}(x, D_{x})\p_{x_{k}}\rho^{\kappa}_{X}),
\]
we see using again (\ref{urlot}) that
\beq\label{arl3}
\lim_{\kappa\to \infty}V_{2}^{\kappa}(X)- \tilde{V}_{2}^{\kappa}(X)= V_{2}^{\infty}(X)\hbox{ exists uniformly for }X\in \rr^{3}
\eeq
and is a bounded continuous function.  The potential $\tilde{V}_{2}^{\kappa}(X)$ can be explicitely evaluated. In fact
\beq\label{arl4}
\begin{array}{rl}
&(\rho^{\kappa}_{X}| c_{X}(x, D_{x})\rho^{\kappa}_{X})\\[2mm]
=&(2\pi)^{-3}\int\e^{\i (x- X)\cdot \xi}c_{X}(x, \xi)\rho^{\kappa}_{X}(x) \hat{\rho}(\kappa^{-1}\xi)\d x \d \xi\\[2mm]
=&(2\pi)^{-3}\int\e^{\i (x- X)\cdot \xi}c_{X}(X, \xi) )\rho^{\kappa}_{X}(x)\hat{\rho}(\kappa^{-1}\xi)\d x \d \xi +O(\kappa^{-1})\log (\kappa)\\[2mm]
=&(2\pi)^{-3}\int c_{X}(X, \xi) |\hat{\rho}|^{2}(\kappa^{-1}\xi) \d \xi +O(\kappa^{-1})\log (\kappa).
\end{array}
\eeq
Similarly
\beq\label{arl2}
\begin{array}{rl}
&(\p_{x_{j}}\rho^{\kappa}_{X}| d_{X}(x, D_{x})\p_{x_{k}}\rho^{\kappa}_{X})\\[2mm]
=&(2\pi)^{-3}\int\e^{\i (x- X)\cdot \xi}\p_{j}\rho^{\kappa}_{X}(x)d_{X}(x, \xi)\i \xi_{k}\hat{\rho}(\kappa^{-1}\xi)\d x \d \xi\\[2mm]
=&(2\pi)^{-3}\int\e^{\i (x- X)\cdot \xi}\rho^{\kappa}_{X}(x)d_{X}(x, \xi)\xi_{j}\xi_{k}\hat{\rho}(\kappa^{-1}\xi)\d x \d \xi\\[2mm]
&-(2\pi)^{-3}\int\e^{\i (x- X)\cdot \xi}\rho^{\kappa}_{X}(x)\p_{j}d_{X}(x, \xi)\i \xi_{k}\hat{\rho}(\kappa^{-1}\xi)\d x \d \xi.
\end{array}
\eeq
The second term in the rhs has  a finite limit when $\kappa\to \infty$. By the same argument as above, we have
\beq\label{arl5}
\begin{array}{rl}
&(2\pi)^{-3}\int\e^{\i (x- X)\cdot \xi}\rho^{\kappa}_{X}(x)d_{X}(x, \xi)\xi_{j}\xi_{k}\hat{\rho}(\kappa^{-1}\xi)\d x \d \xi\\[2mm]
=&(2\pi)^{-3}\int\e^{\i (x- X)\cdot \xi}\rho^{\kappa}_{X}(x)d_{X}(X, \xi)\xi_{j}\xi_{k}\hat{\rho}(\kappa^{-1}\xi)\d x \d \xi+
O(\kappa^{-1}\log(\kappa))\\[2mm]
=&(2\pi)^{-3}\int d_{X}(X, \xi)\xi_{j}\xi_{k}|\hat{\rho}|^{2}(\kappa^{-1}\xi) \d \xi+
O(\kappa^{-1}\log(\kappa)).
\end{array}
\eeq
Using  the definition of $c_{X}(x, \xi)$ and $d_{X}(x, \xi)$ in (\ref{alr1}), we get  that
\[
\begin{array}{rl}
&-c_{X}(X, \xi)+ \12 \sum_{jk}A_{jk}(X)\xi_{j}\xi_{k}d_{X}(X, \xi)\\[2mm]
=&-\12 (h_{0}(X, \xi)+1)^{-\12}K(X, \xi)(K(X, \xi)+1)^{-2}.
\end{array}
\]
Using the definition of $E^{\kappa}(X)$ and (\ref{arl4}), (\ref{arl2}) and (\ref{arl5})
it follows that
\[
\lim_{\kappa\to \infty}\tilde{V}_{2}^{\kappa}(X)- E^{\kappa}(X)\hbox{ exists uniformly for }X\in \rr^{3}.
\]
This completes the proof of the lemma. \qed

\begin{proposition}\label{3.8}
Let
\[
\hat{H}^{\kappa}=U^{\kappa}H^{\kappa}U^{\kappa*}- E^{\kappa}(X).
\]
Then  there exists a bounded below selfadjoint
operator $\hat{H}^{\infty}$ such that
\ben
\item $\hat{H}^{\kappa}$
converges to $\hat{H}^{\infty}$ in norm resolvent sense;
\item
$D(|\hat{H}^{\infty}|^{\12})= D(H_{0}^{\12})$.
\een
\end{proposition}
\proof The proof is analogous to the one in \cite{Ne}, using Theorem
\ref{a.5} so we will only sketch it.
The important point is the convergence of $R^{\kappa}$ as
quadratic form on $D(|H_{0}|^{\12})$ when $\kappa\to \infty$.
The various terms in $R^{\kappa}$ are estimated with the help of
Lemma \ref{a.6}, applied to the coupling operator
$v^{\kappa}=\nabla_{X_{j}}\beta^{\kappa}_{X}$.
From Lemma \ref{3.2} (3), we obtain that
$\omega^{-\alpha}\nabla_{X_{j}}\beta^{\kappa}_{X}$ converges in
$B(\cK, \cK \otimes\ch)$ when $\kappa\to \infty$. The only remaining
point to consider is the fact that powers of the number operator $N$
appear in Lemma \ref{a.6}. This is sufficient in the massive case
since $H_{0}$ dominates $N$. In the massless case, we use the fact
that $\beta^{\kappa}_{X}= F(\omega\geq \sigma/2)\beta^{\kappa}_{X}$.
Therefore if we apply Lemma \ref{a.6}, we can replace $N$ by
$\d\Gamma(\one_{[\sigma/2, +\infty[}(\omega))$, which is dominated by
$H_{0}$. The rest of the proof is standard. \qed

\appendix
\section{Background on pseudodifferential calculus}\label{app1}\init
In this section we recall various standard results on
pseudodifferential calculus that will be needed in the sequel. It is
convenient to use the language of the Weyl-H\"{o}rmander calculus.
\subsection{Symbol classes}\label{app1.1}
We start by recalling the definition of  symbol classes and weights. Let $g$ be a
Riemannian metric on $\rr^{d}$, i.e. a map
\[
g:  \rr^{d}\ni X\mapsto g_{X},
\]
with values in positive definite quadratic forms on $\rr^{d}$. If
$M:
\rr^{d}\to ]0, +\infty[$ is a strictly positive function called a {\em
weight}, one denotes by $S(M, g)$ the symbol class  of functions in
$C^{\infty}(\rr^{d})$ such that
\[
|\prod_{i=1}^{k} (v_{i}\cdot \nabla_{X}) a(X)|\leq
C_{k}M(X)\prod_{i=1}^{k}|g_{X}(v_{i})|^{\12},
\]
uniformly for $X\in \rr^{d}$, $v_{1}, \dots, v_{k}\in \rr^{d}$ and
$k\in \nn$. The  best constants $C_{k}$  are seminorms on $S(M ,g)$.

Usually  $d= 2n$ and one sets $\rr^{d}\ni X=(x, \xi)\in \rr^{n}\times
\rr^{n}$. If
\beq\label{e.app0}
g_{X}=  dx^{2}+ \langle \xi\rangle^{-2} d \xi^{2}
\eeq
and $M(X)= \langle \xi\rangle^{m}$, the symbol class $S(M, g)$ is the
usual symbol class
\[
S^{m}_{1, 0}=\{a\   \ |\p_{x}^{\alpha}\p_{\xi}^{\beta}a(x, \xi)|\leq
C_{\alpha, \beta}\langle \xi\rangle^{m-|\beta|}, \ \alpha, \beta\in
\nn^{n}\}.
\]
For simplicity we will also denote by $S^{p}(\rr)$, $p\in \rr$, the space
\beq\label{e.app1}
S^{p}(\rr)= \{f\   \ |f^{(k)}(\lambda)|\leq C_{k}\langle
\lambda\rangle^{p-k}, \ k \in \nn\},
\eeq
ie $S^{p}(\rr)= S(\langle \lambda\rangle^{p}, \langle
\lambda\rangle^{-2}\d \lambda^{2})$.

If one equips $\rr^{2n}$ with the usual symplectic form
$\sigma$, one can consider  the dual metric $g^{\sigma}_{X}$.
Diagonalising $g_{X}$ in (linear) symplectic coordinates on
$\rr^{2n}$, one can write
\[
g_{X}(dx, d\xi)= \sum_{i=1}^{n}\frac{d x_{i}^{2}}{a_{i}^{2}(X)}+
\frac{d \xi_{i}^{2}}{\alpha_{i}^{2}(X)},
\]
and
\[
g_{X}^{\sigma}(dx, d\xi)= \sum_{i=1}^{n} \alpha_{i}^{2}(X)dx_{i}^{2}+
a_{i}^{2}(X)d\xi_{i}^{2}.
\]
One introduces also the two functions $\lambda(X)$, $\Lambda(X)$ which
are the best functions such that
\[
\lambda(X)^{2}g_{X}\leq g^{\sigma}_{X}\leq \Lambda(X)^{2}g_{X},
\]
equal to
\[
\lambda(X)=\min_{i}a_{i}(X)\alpha_{i}(X), \ \Lambda(X)=
\max_{i}a_{i}(X)\alpha_{i}(X).
\]
The function $\lambda(X)$ plays the role of the Planck constant.

One says that $g$ is a {\em H\"{o}rmander metric}, if the following
conditions are satisfied
\ben
\item {\em uncertainty principle: } $\lambda(X)\geq 1$;
\item {\em slowness: } there exists $C>0$ such that
\beq\label{e.a1}
g_{Y}(X-Y)\leq C^{-1}\ \Rightarrow \
\left(g_{Y}(\cdot)/g_{X}(\cdot)\right)^{\pm 1}\leq C;
\eeq
\item {\em temperateness: } there exist $C>0$, $N\in \nn$ such that
\beq\label{e.a2}
\left(g_{Y}(\cdot)/g_{X}(\cdot)\right)^{\pm 1}\leq C\left(1 +
g^{\sigma}_{Y}(Y-X)\right)^{N}.
\eeq
\een
One says that a weight $M$ is {\em admissible} for $g$ if there exist
$C>0$, $N\in \nn$ such that
\beq\label{e.a3}
\left(M(Y)/M(X)\right)^{\pm 1}\leq\left\{
\begin{array}{l}
C,\hbox{ for }g_{Y}(X-Y)\leq C^{-1},\\
C(1+ g^{\sigma}_{Y}(X-Y))^{N}, \hbox{ for }X, Y\in \rr^{2n}.
\end{array}
\right.
\eeq
The metric $g$ is {\em geodesically temperate} if $g$ is temperate
and if there exist $C>0$ and $N\in \nn$ such that
\beq\label{e.a4}
\left(g_{Y}(\cdot)/g_{X}(\cdot)\right)^{\pm 1}\leq C(1+ d^{\sigma}(X,
Y))^{N},
\eeq
where $d^{\sigma}$ is the geodesic distance for  the metric
$g^{\sigma}$.

The metric $g$ is {\em strongly slow} if there exists $C>0$ such that
\beq\label{e.a5}
g^{\sigma}_{Y}(X- Y)\leq C^{-1}\Lambda(Y)^{2}\ \Rightarrow \
\left(g_{Y}(\cdot)/g_{X}(\cdot)\right)^{\pm 1}\leq C.
\eeq

\begin{lemma}\label{a.1}
The metric $dx^{2}+ \langle \xi\rangle^{-2}d \xi^{2}$ and weight
$\langle \xi\rangle^{\alpha}$ for $\alpha\in \rr$ satisfy all the
above conditions.
\end{lemma}
\proof Most conditions are immediate, except the last two. To check
(\ref{e.a4}), we note that $d^{\sigma}(X, Y)\leq
|\xi- \eta|$, from which (\ref{e.a4})  follows. (\ref{e.a5}) follows
from the fact that $\Lambda(X)= \langle \xi\rangle$.\qed
\begin{lemma}\label{a.2}
Assume that  $(g_{i},M_{i})$, $i=1, 2$ are two metrics and weights on
$\rr^{2n_{i}}$ satisfying all the above conditions. Then $(g, M)$ on
$\rr^{2n}$ satisfy all the above conditions for $n= n_{1}+ n_{2}$ and
\[
g_{X}(dx)= g_{X_{1}}(dx_{1})+ g_{X_{2}}(dx_{2}), \ M(X)= M_{1}(X_{1})+
M_{2}(X_{2}).
\]
\end{lemma}
\subsection{Pseudodifferential calculus}\label{app1.2}
To a symbol $a\in S'(\rr^{2n})$, one can associate the operator
defined by
\beq\label{e.a10}
a^{\rm w}(x, D)u(x)=(2\pi)^{-n}\int\e^{\i
(x-y)\cdot \xi}a(\frac{x+y}{2},\xi)u(y)dyd\xi,
\eeq
called the {\em Weyl quantization} of $a$,
which is well defined as a bounded operator from $S(\rr^{n})$ into
$S'(\rr^{n})$.  Let $(g, M)$ be a metric and weight satisfying
(\ref{e.a1}), (\ref{e.a2}), (\ref{e.a3}). We set
\[
\Psi^{\rm w}(M, g)= \{a^{\rm w} \   \ a\in S(M, g)\}.
\]

If $a\in S(M, g)$ then $a^{\rm w}$ sends
$S(\rr^{n})$ into itself. Moreover as quadratic forms on $S(\rr^{n})$
\[
(a^{\rm w})^{*}= \overline{a}^{\rm w}.
\]
One often uses also  the $(1, 0)$
quantization defined by
\beq\label{zorb}
 a^{\rm 1,0}(x, D)u(x)=(2\pi)^{-n}\int\e^{\i
(x-y)\cdot \xi}a(x, \xi)u(y)dyd\xi.
\eeq
One has with obvious notations
\beq\label{e.a6bis}
\Psi^{\rm w}(M, g)= \Psi^{(1, 0)}(M, g).
\eeq
Moreover
\beq\label{e.a6ter}
\Psi^{\t}(M_{1}, g)\times \Psi^{\t}(M_{2}, g)\subset
\Psi^{\t}(M_{1}M_{2}, g),
\eeq
where $\t={\rm w}$ or $(1, 0)$ and if $a\in S(M, g)$
\beq\label{e.a6quater}
a^{\rm w}(x, D_{x})= b^{\rm (1, 0)}(x, D_{x}), \hbox{ where }a-b\in
S(M\lambda^{-1}, g).
\eeq

Let now $g$ be the standard metric defined in (\ref{e.app0}) and
$H^{s}(\rr^{d})$  be the Sobolev space of order $s\in \rr$. Then
\beq\label{e.app00}
\Psi^{\t}(\langle \xi\rangle^{p}, g)\subset B(H^{s}(\rr^{d}),
H^{s-p}(\rr^{d})),
\eeq
and the norm of $a^{\t}$ in $B(H^{s}(\rr^{d}),
H^{s-p}(\rr^{d}))$ is controlled by a finite number of seminorms of
$a$ in $S(\langle \xi\rangle^{p}, g)$.

\subsection{Functional calculus for pseudodifferential
operators}\label{qpp.3}
Assume that the weight $M$ satisfies
\beq
M(X)\leq C(1+ \lambda(X))^{N}, \ C>0, \ N\in \nn.
\label{e.a6}
\eeq
A symbol $a\in S(M, g)$ is {\em elliptic} if
\beq
 1+ |a(X)|\geq C^{-1}M(X).
\label{e.a7}
\eeq
The following theorem is shown in \cite[Cor. 4.5 ]{Bo}
\bet\label{a.3}
Assume that $(M, g)$ satisfy all the conditions in Subsect.
\ref{app1.1}. Assume moreover that $M\geq 1$, $a\in S(M, g)$ is real
and elliptic, and $a^{\rm w}$ is essentially selfadjoint on
$S(\rr^{n})$.  Then  if $f\in S^{p}(\rr)$, the operator $f(a^{\rm w})$ belongs
to $\Psi^{\rm w}( M^{p}, g)$.
\eet
The following result can easily be obtained.
\begin{proposition}\label{a.4}
Assume the hypotheses of Theorem  \ref{a.3}. Then
\[
f(a^{\rm w})- f(a)^{\rm w}\in \Psi^{\rm w}( M^{p}\lambda^{-1}, g),
\]
where the function $\lambda(X)$ is defined in Subsect. \ref{app1.1}.
\end{proposition}
Note that the same result holds for the $(1, 0)$  quantization, thanks
to (\ref{e.a6quater}).

\proof one first proves the result for $f(\lambda)= (\lambda-z)^{-1}$,
$z\in \cc\backslash\rr$,
which amounts to construct a so-called parametrix for $a^{\rm w}-z$.
From symbolic calculus it follows that if $b_{z}(x, \xi)= (a(x,
\xi)-z)^{-1}$, then $b_{z}^{\rm w}(a^{\rm w}-z)-\one\in \Psi^{\rm
w}( \lambda^{-1}, g)$. To extend the result to arbitrary functions one
expresses $f(a^{\rm w})$  in terms of $(a^{\rm w}-z)^{-1}$ using the
well known
functional calculus formula based on an almost analytic extension of
$f$ (see eg \cite[Prop. C.2.2]{DG}). \qed

\subsection{Various estimates}
\def\bb{B(\cK, \cK\otimes\ch)}
The following lemma is proved in \cite[Lemma 3.3]{Am}.
\begin{lemma}\label{a.6}
For  $s \in [0,1] \;$ , and $ v_{i} \in B(\cK, \cK\otimes\ch),\;i=1,2$ we have
\[
\begin{array}{l}
1)\   \|(N+1)^{-\frac{s}{2}} \,a(v_{1})\,
(H_{0}+1)^{-\frac{1-s}{2}}\| \,\leq  \,\|\omega ^{\frac{s-1}{2}}
\,v_{1} \|_{\bb},\\[2mm]
2)\ \|(H_{0}+1)^{-\frac{s}{2}}
\, a^{*}(v_{1}) \, (N+1)^{-\frac{1-s}{2}}\| \,\leq  \,\|\omega
^{-\frac{s}{2}} \, v_{1}  \|_{\bb},\\[2mm]
3) \ \|
(N+1)^{-s} \, a( v_{1}) \, a(v_{2}) \, (H_{0}+1)^{-1+s} \| \leq
 \,\|\omega ^{-\frac{1-s}{2}} \,v_{1} \|_{\bb}
\, \|\omega ^{-\frac{1-s}{2}} \,v_{2} \|_{\bb},\\[2mm]
4) \ \|(H_{0}+1)^{-s} \, a^{*}(v_{1}) a^{*}(v_{2}) (N+1)^{-1+s}\|
\leq
 \,\|\omega ^{-\frac{s}{2}} \,v_{1} \|_{\bb}
\, \|\omega ^{-\frac{s}{2}} \,v_{2} \|_{\bb}.
\end{array}
\]
\end{lemma}
The following theorem follows from the KLMN theorem and \cite[Theorem  VIII.25]{RS1}.
\bet\label{a.5}
Let $H_{0}$ be a positive selfadjoint operator on  a Hilbert space ${\cal {H}}$. Let for
$\kappa <\infty$, $B_{\kappa}$  be quadratic forms on $D(H_{0}^{\frac{1}{2}})$ such that
\[
|B_{\kappa}(\psi,\psi)| \leq a \, ||H_{0}^{\frac{1}{2}} \psi||^{2}
+ b \, ||\psi||^{2},
\]
where $a <1$ uniformly in $\kappa$ and $B_{\kappa} \rightarrow
B_{\infty}$ on $D(H_{0}^{\frac{1}{2}})$.\\
Then
\ben
\item there exists a selfadjoint operator $H_{\kappa}$ with
$D(H_{\kappa}) \subset D(H_{0}^{\frac{1}{2}})$ and
\[
(H_{\kappa} \psi, \psi)=B_{\kappa}(\psi,\psi)+(H_{0}^{\frac{1}{2}} \psi,
H_{0}^{\frac{1}{2}} \psi) , \; \psi \in D(H_{\kappa}) {\rm \;for \;}
\kappa \leq \infty.
\]
\item the resolvent $(z-H_{\kappa})^{-1}$ converges in norm to
$(z-H_{\infty})^{-1}$.
\item $\e^{it H_{\kappa}}$ converges strongly to $\e^{it H_{\infty}}$  when
$\kappa\to +\infty$.
\een
\eet

\end{document}